\documentclass[%
pre,
 amsmath,amssymb,
preprint,
]{revtex4-1}

\usepackage{graphicx}
\usepackage{dcolumn}
\usepackage{bm}
\usepackage{gensymb}
\usepackage[ruled,vlined]{algorithm2e}
\usepackage{algpseudocode}
\usepackage{algcompatible}
\usepackage{newfloat}
\usepackage{color, soul}

\usepackage[russian,english]{babel}

\begin{document}
\title{Anomalous particle diffusion influenced by angular heterogeneity}

\author{Kejie Chen}
 \affiliation{Department of Mechanical Engineering, University of Michigan, Ann Arbor, MI 48105, USA }
\author{Bogdan I. Epureanu}
 \altaffiliation{Corresponding author: epureanu@umich.edu }
 \affiliation{Department of Mechanical Engineering, University of Michigan, Ann Arbor, MI 48105, USA }

\begin{abstract}
A generalized persistent random walk (GPRW) model to study anomalous particle diffusion influenced by angular heterogeneity is presented. Consider the motion of a particle is composed of many consecutive straight line segments.  At the end of each straight motion, the particle can switch to a new moving direction. Angular heterogeneity occurs when the probability of choosing a moving direction is non-uniformly distributed and influenced by surrounding environment and by the particle past conditions. Based on the model, we show that particles perform Gaussian Fickian diffusion with directional drifting in a memoryless system with spatial dependent angular heterogeneity. When the probability of maintaining the current moving direction is correlated with the particle speed, these particles perform Fickian diffusion, but their full distribution diverges from Gaussian. When the probability of maintaining te current moving direction is correlated with the historical moving distance, particles perform superdiffusion. The GPRW framework presented can be applied to study protein diffusion, cell migration and solute transport in complex heterogeneous environments.

\end{abstract}

\maketitle


\section{Introduction}
Anomalous diffusion and transport phenomena other than Gaussian and Fickian diffusion have been widely observed in complex systems, such as particle diffusion in turbulent flows and ionic solvent, transport in porous media, bacteria and cell migration, and spreading of epidemics \cite{RichardsonL, BenjaminR, RheeI, ArielG, Zhonghan}. The non-Gaussian and nonlinear diffusion dynamics are usually correlated with important functions and properties. For example, the superdiffusive migratory dynamics of predators improves the chance of catching a rare prey \cite{DavidWS}. The non-Gaussian and non-Fickian transport of contaminant plumes has a major impact on water resource management \cite{BottacinB, DanielB}. Therefore, dissecting the underlying causes of different types of anomalous diffusion and transport dynamics could provide useful insights as well as help designing and optimizing system characteristics. To understand the anomalous diffusion and transport dynamics, mathematical and statistical physics approaches have been developed to link microscopic particle motions with macroscopic patterns and distributions. Specifically, random walk approaches are advantageous because they provide flexibility of model construction and explicit physical meaning of model variables and parameters. They have been widely applied to study and understand phenomena involved in cell migration, stock price fluctuations and solute transport in porous beds \cite{PeiHsun, EugeneF, Delgoshaie}. For example, Boano et al. applied the continuous time random walk model to study the hyporheic exchange and in-stream storage influenced by bedforms with different sizes and local heterogeneity \cite{BoanoF}.

Among random walk approaches, the most well-known is the diffusion equation derived based on heat conduction by Fourier more than a century ago \cite{FourierB}. The diffusion equation describes Gaussian and Fickian motion of a particle where its mean square displacement (MSD, denoted as $\langle \boldsymbol{x}^2 \rangle$) increases linearly in time, $\langle \boldsymbol{x}^2 \rangle = 4Dt$, where $D$ is the diffusion coefficient. However, a non-physical drawback of the diffusion equation is the possibility of having infinite particle speed. In addition, experiments show that the MSD can be nonlinearly varying in time. Specifically, $\langle \boldsymbol{x}^2 \rangle = 4D_{\alpha}t^{\alpha}$. When $\alpha < 1$, the particle performs subdiffusion. When $1<\alpha<2$, the particle performs superdiffusion. To consider only finite speeds for the moving particles, a persistent random walk (PRW) model is proposed. In a two dimensional (2D) space, the PRW model considers the motion of  a particle as composed of many consecutive straight line segments. Specifically, a particle starts from a point $0$ and moves in a $\theta_1$ direction with a constant speed $U$ for time $\tau_{p1}$. Afterward, the particle switches direction and moves in a $\theta_2$ direction with speed $U$ for a time duration $\tau_{p2}$. The particle repeats the moving and turning process $n$ times. The moving direction, $\theta_i, \, i \in [1, \, n]$, are uniformly distributed between 0 and $2\pi$. The PRW model describes the probability of finding a particle at a distance between $r$ and $r+dr$ after $n$ moving and turning events starting from point $0$. When the time duration of all events is the same ($\tau_{pi} = \tau_{pj}, \, \forall \, i, j \in [1, n]$), the PRW model can be simplified into the telegraph equation, which captures the transient ballistic motion at short times and the Gaussian Fickian diffusion at long times \cite{WeissG}. When the particle moves under random external perturbations, the duration of a moving and turning event is random. When the distribution of event duration follows an exponential distribution, the PRW model can be simplified into a Fokker Planck equation \cite{RaziNaqvi}. When the event duration follows a long-tailed distribution, the particle motion is superdiffusive. To consider the subdiffusive dynamics where particles spread slower than in Fickian diffusion, a long-tailed distribution of waiting time between consecutive events is added to the model. Fractional telegraph equations and fractional Fokker Planck equations can further be derived from the PRW model for superdiffusive or subdiffusive motions \cite{GoldsteinJA, KlafterJ}. To consider fluctuations in the particle speed, Zaburdaev et al. \cite{ZaburdaevV2} proposed a random walks with random velocity model. Based on the model, they discovered ballistic hump-like regions in the particle distribution. While most previous work focuses on particle moving time and speed, the effects of angular heterogeneity on diffusion and transport dynamics is largely unknown.

In this work, we extend the PRW framework to generic situations, and proposed a generalized persistent random walk (GPRW) model which can be used to study the influence of angular heterogeneity on particle diffusion and transport in biological systems, including intracellular protein transport and cell migration. We consider that the moving direction of a particle is spatial dependent, correlated with particle speed, or correlated with particle moving history. Based on the GPRW model, the time series of particle distributions and MSDs are calculated for three types of angular heterogeneity. Specifically, in the first situation, the probability of choosing a moving direction depends on particle location but is independent of particle historical motion. In this case, we show that the GPRW model can be simplified into a partial differentiation equation which implies that the temporal rate of change of particle density in phase space is influenced by the spatial and angular particle flux. At long times, the particle motion can be decomposed into Fickian diffusion and directional drifting. In the second situation, the probability of maintaining the same moving direction is correlated to the particle speed, namely, when the particle moves fast, the particle maintains the moving direction for a long time. In this case, the particle performs Fickian diffusion with a significantly improved diffusion coefficient, and its distribution diverges from Gaussian. In the third situation, the probability of maintaining the same moving direction is reinforced by the time the particle has already traveled in that same direction. In this case, the particle motion is shown to be superdiffusive. In addition, the implications of the modeling results to biological systems are discussed.

\begin{figure}[h]
	\begin{center}
		\includegraphics[width=0.9\textwidth]{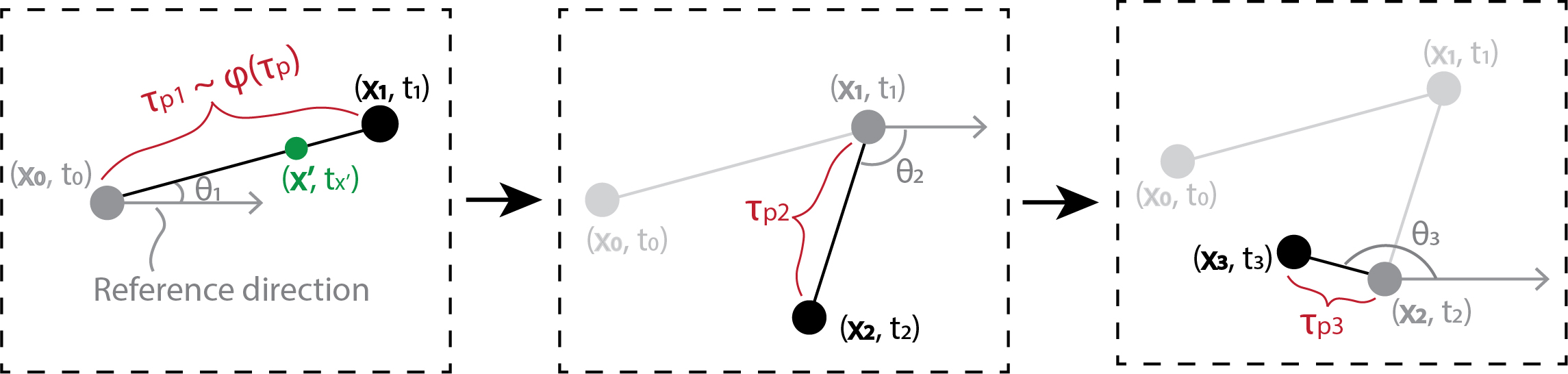}
		\caption{\textbf{Schematic illustration of the 2D microscopic motion of a particle.} Figures in three dashed box represent three snapshots of particle motion when the particle is at the end of a sojourn and will switch moving direction in the next time step. Black dots represent current locations of the particle and grey dots represent historical particle locations.}
	\end{center}
\end{figure}
\section{GPRW model framework}
Consider a particle motion modeled as a series of connecting straight line segments as in the PRW model \cite{MasoliverJ, MasoliverJ2}. Each line segment is referred to as a sojourn. For example, in a 2D situation (Figure 1), the motion from $(\boldsymbol{x}_0, \, t_0)$ to $(\boldsymbol{x}_1, \, t_1)$ is a sojourn. The particle velocity $\boldsymbol{u}$ at position $\boldsymbol{x}$ and time $t$ is a 2D vector, which is characterized by a magnitude $U$ and a direction $\theta$. $\boldsymbol{u}$ is assumed to be the same during a sojourn. At the end of a sojourn, the direction $\theta$ and magnitude $U$ can change. For example, in Figure 1, $\boldsymbol{u}_{\boldsymbol{x}', \, t_{x'}} = \boldsymbol{u}_{\boldsymbol{x}_0, \, t_0} \neq \boldsymbol{u}_{\boldsymbol{x}_1, \, t_1}$. The duration of a sojourn is referred to as persistence (or persistent time), denoted as $\tau_p$. For example, $\tau_{p1} = t_1 - t_0$ is the persistence of a sojourn from $(\boldsymbol{x}_0, \, t_0)$ to $(\boldsymbol{x}_1, \, t_1)$. The distribution of $\tau_p$ is $\varphi(\tau_p)$, and the cumulative distribution is $\Phi (\tau_p) = \int_{\tau_p}^{\infty} \varphi (t) dt$. At the end of a sojourn, the probability of switching to velocity $\boldsymbol{u}$ is denoted as $r(\boldsymbol{u}; \, \chi)$, where $\chi$ represents influential factors such as particle position and historical motion. For example, $r(\boldsymbol{u}_1; \, \boldsymbol{u}_0, \, \boldsymbol{x}_1, \, t_1)$ represents the probability of switching to velocity $\boldsymbol{u}_1$, which is influenced by particle location $\boldsymbol{x}_1$, time $t_1$ and previous velocity $\boldsymbol{u}_0$. The particle distribution in phase space is described by a probability density $p(\boldsymbol{u}, \boldsymbol{x}, t)$, which is the probability of observing a particle at position $\boldsymbol{x}$ at time $t$ with velocity $\boldsymbol{u}$. Thus, $\int_{-\boldsymbol{\infty}}^{\boldsymbol{\infty}} \, \int_{-\boldsymbol{\infty}}^{\boldsymbol{\infty}} \, p(\boldsymbol{u}, \boldsymbol{x}, t) d \boldsymbol{u} \, d \boldsymbol{x} = 1$. To calculate $p(\boldsymbol{u}, \boldsymbol{x}, t)$, define a rate $q(\boldsymbol{u}, \boldsymbol{x}, t)$ which represents the probability that particles whose sojourns end at position $\boldsymbol{x}$ and time $t$ will move with a new velocity $\boldsymbol{u}$. Note that $\int_{-\boldsymbol{\infty}}^{\boldsymbol{\infty}} \, \int_{-\boldsymbol{\infty}}^{\boldsymbol{\infty}} \, q(\boldsymbol{u}, \boldsymbol{x}, t) d \boldsymbol{u} \, d \boldsymbol{x} \neq 1$. Note that, in this work, we define two new notations $\int_{-\boldsymbol{\infty}}^{\boldsymbol{\infty}} \boldsymbol{\cdotp} d \boldsymbol{u}$ and $\int_{-\boldsymbol{\infty}}^{\boldsymbol{\infty}} \boldsymbol{\cdotp} d \boldsymbol{x}$ for a 2D domain of coordinates $u_1$, $u_2$ and $x_1$ and $x_2$. Specifically, $\int_{-\boldsymbol{\infty}}^{\boldsymbol{\infty}} \boldsymbol{\cdotp} d \boldsymbol{u} = \int_{u_1=-\infty}^{u_1=\infty} \int_{u_2=-\infty}^{u_2=\infty} \boldsymbol{\cdotp} d u_1 \, du_2$, and $\int_{-\boldsymbol{\infty}}^{\boldsymbol{\infty}} \boldsymbol{\cdotp} d \boldsymbol{x} = \int_{x_1=-\infty}^{x_1=\infty} \int_{x_2=-\infty}^{x_2=\infty} \boldsymbol{\cdotp} d x_1 \, d x_2$. 

Inspired by the PRW model \cite{MasoliverJ2}, $p(\boldsymbol{u}, \boldsymbol{x}, t)$ and $q(\boldsymbol{u}, \boldsymbol{x}, t)$ satisfy the probability master equations
\begin{align}
\begin{split}
q(\boldsymbol{u},\boldsymbol{x},t) & =  p_0 (\boldsymbol{u}, \boldsymbol{x}) \delta (t) + \\
& \int_{-\boldsymbol{\infty}}^{\boldsymbol{\infty}} \int_{-\boldsymbol{\infty}}^{\boldsymbol{\infty}} \int_0^t q(\boldsymbol{u}',\boldsymbol{x}', t') \, \delta \big(\boldsymbol{x}-\boldsymbol{x}'-\boldsymbol{u}' (t-t') \big) \, \varphi (t-t') r(\boldsymbol{u}; \, \chi) \, d \boldsymbol{u}' d \boldsymbol{x}' dt', \\
\end{split}
\\
& p(\boldsymbol{u}, \boldsymbol{x}, t)  = \int_{-\boldsymbol{\infty}}^{\boldsymbol{\infty}} \int_0^t q(\boldsymbol{u}, \boldsymbol{x}', t') \, \delta\big( \boldsymbol{x}-\boldsymbol{x}'-\boldsymbol{u} (t-t') \big) \, \Phi (t-t')  d \boldsymbol{x}' dt',
\end{align}

\noindent where $p_0 (\boldsymbol{u}, \boldsymbol{x})$ is the initial probability distribution. Equation (1) is derived from the consideration that when a particle arrives at position $\boldsymbol{x}$ and at time $t$, it is because at a previous time $t^{'}$ ($0<t^{'}<t$) the particle starts moving from location $\boldsymbol{x}^{'}$ with a velocity $\boldsymbol{u}^{'}$, where $\boldsymbol{x}-\boldsymbol{x}^{'}=\boldsymbol{u}^{'}(t-t^{'})$. After arriving at position $\boldsymbol{x}$ at time $t$, the particle velocity changes to $\boldsymbol{u}$. Equation (2) considers that when a particle passes by location $\boldsymbol{x}$ at time $t$, at a previous time $t^{'}$ ($0<t^{'}<t$) the particle should have started to move from location $\boldsymbol{x}^{'}$ with a velocity $\boldsymbol{u}$, where $\boldsymbol{x}-\boldsymbol{x}^{'}=\boldsymbol{u}(t-t^{'})$, and the persistent time $\tau_p$ of the particle is equal to or longer than $t-t^{'}$. It should be noted that $q(\boldsymbol{u}, \boldsymbol{x}, t)$ only represents particles at the end of a sojourn, while $p(\boldsymbol{u}, \boldsymbol{x}, t)$ represents particles along a sojourn and at the end of a sojourn.

Based on the GPRW model, the effects of three types of heterogeneity on particle distribution and dynamics are investigated. In the following three sections, we first introduce the type of heterogeneity inspired by real world phenomena. Then, the modeling approach and results based on GPRW framework are discussed. Implications and potential applications in biological systems, including intracellular protein diffusion and cell migration, are discussed at the end of each section. 

\section{Fickian diffusion and drifting}
Many chemical and biological systems, such as colloidal suspensions and polymer networks, are driven only by thermal fluctuations and contain white and uncorrelated systematic noise. These systems are memoryless, which means particles moving inside the system are not influenced by historical conditions. However, properties of the system, such as the density and polarity, can be spatially heterogeneous. For example, a colloidal suspension with spatial varying density can be memoryless but heterogeneous. To study the dynamics of particles diffusing in such media, master equations (1) and (2) are simplified using $r(\boldsymbol{u}; \, \chi) = r(\boldsymbol{u}; \, \boldsymbol{x})$ and $\varphi (\tau_p) = \gamma e^{-\gamma \tau_p}$, where $\gamma$ is a model parameter. $\gamma^{-1}$ is the mean persistent time, which is also the time scale of the system. Specifically, in a memoryless system, the persistent time follows an exponential distribution. The angular heterogeneity, described by the probability of switching to velocity $\boldsymbol{u}$, depends only on the particle location $\boldsymbol{x}$. Equations (1) and (2) can be rewritten as 
\begin{align}
\begin{split}
q(\boldsymbol{u},\boldsymbol{x},t) & =  p_0 (\boldsymbol{u}, \boldsymbol{x}) \delta (t) + \gamma \, r(\boldsymbol{u}; \, \boldsymbol{x})  \int_{-\boldsymbol{\infty}}^{\boldsymbol{\infty}} \int_0^t q \big( \boldsymbol{u}',\boldsymbol{x}-\boldsymbol{u}' (t-t'), t' \big) \, e^{-\gamma(t-t')} \, d \boldsymbol{u}' dt', \\
\end{split}
\\
& p(\boldsymbol{u}, \boldsymbol{x}, t)  = \int_0^t q \big( \boldsymbol{u}, \boldsymbol{x}-\boldsymbol{u} (t-t'), t' \big) \, e^{-\gamma (t-t')}  dt',
\end{align}

By performing a Taylor expansion and a Laplace transform (detailed derivations are provided in Appendix A), Equations (3) and (4) can be simplified to a partial differential equation as
\begin{equation}
\displaystyle (\partial_t + \boldsymbol{u} \cdot \boldsymbol{\bigtriangledown}) p(\boldsymbol{u}, \boldsymbol{x}, t) = \gamma \, \int_{-\boldsymbol{\infty}}^{\boldsymbol{\infty}} r(\boldsymbol{u}; \,\boldsymbol{x}) \, p(\boldsymbol{u}', \boldsymbol{x}, t) d \boldsymbol{u}' - \gamma p(\boldsymbol{u}, \boldsymbol{x}, t).
\end{equation}

Equation (5) states that the rate of change in the probability that a particle that moves with velocity $\boldsymbol{u}$ at position $\boldsymbol{x}$ is determined by its spatial motion (particle moves into or out of position $\boldsymbol{x}$) and the motion in phase space (particle changes velocity between $\boldsymbol{u}$ and $\boldsymbol{u}'$). Note that a similar equation is derived in \cite{Oleksandr} based on the Langevin approach. 

When the magnitude of the velocity $U$ is a constant, the dynamics only depends on the direction of $\boldsymbol{u}$. Thus, $r(\boldsymbol{u}; \, \boldsymbol{x}) = r(\theta; \boldsymbol{x})$. By performing a moment expansion on Equation (5) (detailed derivations are provided in Appendix B), at long times, particle motion follows
\begin{equation}
\displaystyle \partial_t \rho (\boldsymbol{x}, t) = \dfrac{U^2}{2\gamma} \Delta \rho(\boldsymbol{x}, t) - U \,  \rho (\boldsymbol{x}, t)\bigtriangledown \cdot \bigg[ \int_0^{2\pi} r(\theta; \boldsymbol{x}) \cos\theta d \theta, \, \int_0^{2\pi} r(\theta; \boldsymbol{x}) \sin \theta d\theta \bigg].
\end{equation}

\noindent where $\rho(\boldsymbol{x}, t) = \int_{-\boldsymbol{\infty}}^{\boldsymbol{\infty}}p(\boldsymbol{u}, \boldsymbol{x}, t)d \boldsymbol{u}$ is the particle distribution in real space. The asymptotic solution (6) implies that the particle motion is a combination of diffusion and directional drifting. Diffusion, described by the first term on the right side, can be represented by a point-wise diffusion coefficient $D=\dfrac{U^2}{2\gamma}$. Directional drifting, described by the second term on the right side, can be represented by a net velocity (advection velocity) $\boldsymbol{u}_{av} = U \bigg[ \int_0^{2\pi} r(\theta; \boldsymbol{x}) \cos \theta d \theta, \, \, \int_0^{2\pi} r(\theta; \boldsymbol{x}) \sin \theta d \theta \bigg]$. 

\begin{figure}[h]
	\begin{center}
		\includegraphics[width=0.8\textwidth]{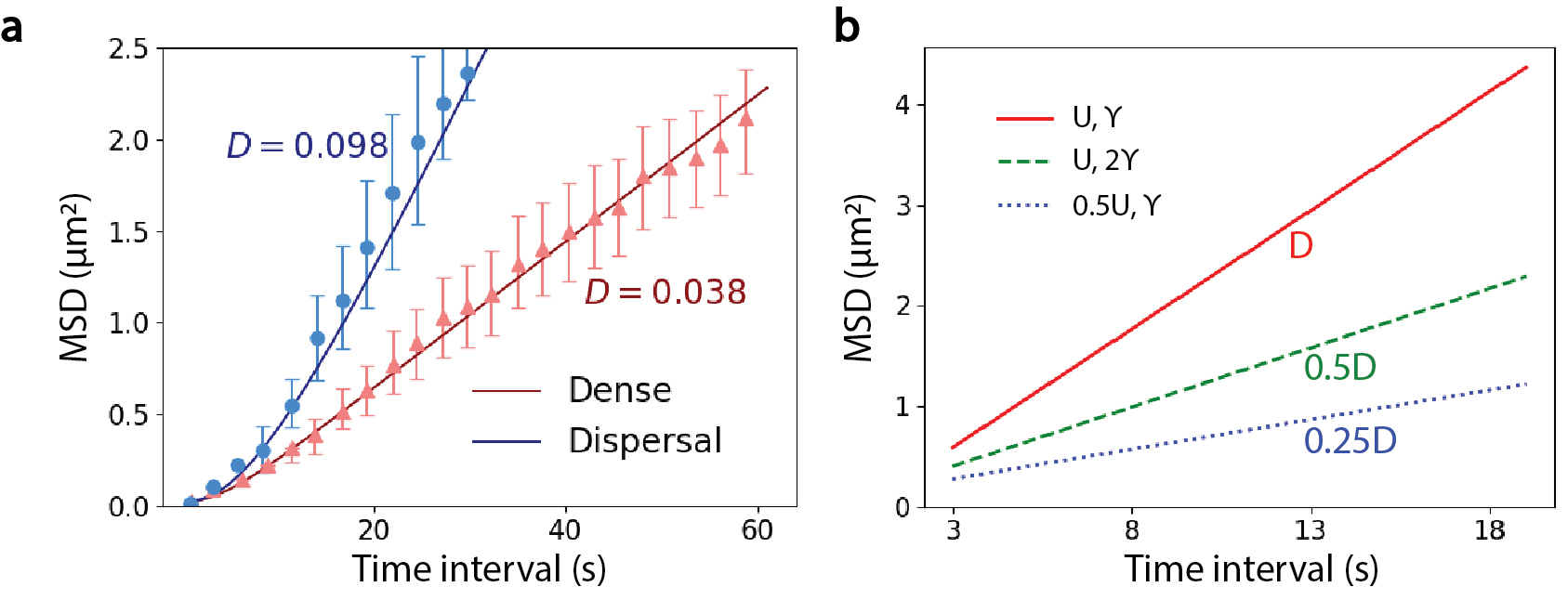}
		\caption{\textbf{MSD of a particle in a static, homogeneous and memoryless environment.} \textbf{a}, MSDs of granules transported by molecular motors in a dense network and a dispersal network of actin filaments. Triangles and dots represent experimental measurements \cite{JosephSn}. Solid lines are the results calculated using Equation (5). \textbf{b}, The diffusion coefficient is determined by the microscopic particle speed $U$ and the mean persistent time $\gamma^{-1}$, for $\varphi(\tau_p)=\gamma e^{-\gamma \tau_p}$.}
	\end{center}
\end{figure}

To verify the model, the MSDs of particles obtained by numerically solving Equation (5) are compared with experimental measurements. Specifically, in the experiment \cite{JosephSn}, granule particles are transported in a dense network and a dispersal network of actin filaments by molecular motors. Inside the network, a granule particle is moved by motors along a straight filament for some distance. At the intersection of filaments or when all motors detach from the filament, the moving direction of the granule changes. The GPRW is applied to capture the microscopic motion of granules. Since the networks of actin filaments are homogeneous. Thus, $r(\boldsymbol{u}; \,  \boldsymbol{x}) = \dfrac{1}{2 \pi}$. Granule speed $U$ is measured as $72 \pm 4 \, nm/s$ in a dense network, and $80 \pm 5 \, nm/s$ in a dispersal network. The measured mean particle persistence $\gamma^{-1}$ is $2.8 \, s$ in a dense network and $7.5 \, s$ in a dispersal network. The MSDs calculated by Equation (5) agrees well with the experimental measurements at both short and long times, as shown in Figure 2 a. Specifically, at the initial transient stage ($t \leq \gamma^{-1}$), the particle motion is superdiffusive.  At long times ($t \geq \gamma^{-1}$), the granule particle performs Fickian diffusion with a diffusion coefficient $D = \dfrac{U^2}{2 \gamma}$ (Figure 2 b). 

The GPRW model can be applied to predict protein or cell distribution in heterogeneous environments. In contrast to a homogeneous environment, inside many types of cells, the positive ends of actin filaments are aligned towards the cell periphery. Thus, intracellular particles, such as proteins and organelles, transported on the actin filaments have a large probability to choose a moving direction towards cell periphery. In cell migration, due to chemotaxis and durataxis properties, chemical and rigidity gradients modify the probability distribution of migratory directions of cells \cite{NovikovaE, HuangSui}. As a simple demonstration, consider a case where the probability of switching to moving direction $\theta$ follows three types of distributions (i.e. three types of heterogeneity). Define type 1 heterogeneity as $r(\theta; \, \boldsymbol{x}) =  \dfrac{\cos^4 \theta}{\int_0^{2 \pi} \cos^4 \theta d \theta}$; type 2 heterogeneity as $r(\theta; \, \boldsymbol{x}) = \dfrac{\cos^2 \theta}{\int_0^{2 \pi} \cos^2 \theta d \theta}$ and type 3 heterogeneity as $r(\theta; \, \boldsymbol{x}) = \dfrac{\cos^2 \theta + 0.5}{\int_0^{2 \pi} (\cos^2 \theta + 0.5) d \theta}$. In the three types of heterogeneous systems, the particle has a larger probability to move toward left ($\theta = \pi$) than toward right ($\theta = 0$ or $2 \pi$). Consequently, the MSDs of particles follow power law distributions (MSD $\propto t^{\alpha}, \, \alpha>1$), as shown in Figure 3 a. At long times, as shown in Figure 3 b, the particle motion is diffusive with drifting towards the left. Moreover, since type 3 distribution $r(\theta; \, \boldsymbol{x})$ is closer to a uniform distribution compared with type 1 and 2 distributions, the directional drifting in the system of type 3 heterogeneity is slower, and the power law exponent $\alpha$ of its MSD is closer to 1.

\begin{figure}[h]
	\begin{center}
		\includegraphics[width=0.85\textwidth]{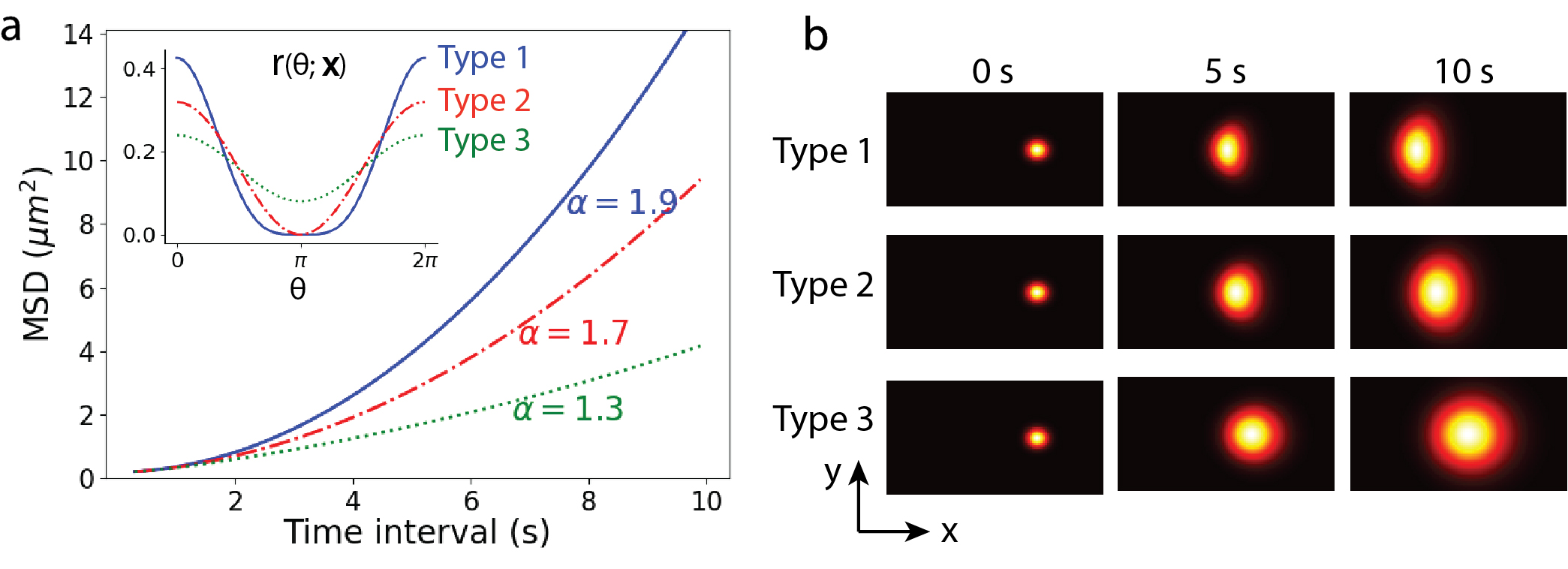}
		\caption{\textbf{The angular heterogeneity leads to the directional drifting of particles.} \textbf{a}, Three types of distributions $r(\theta; \boldsymbol{x})$ and the corresponding MSDs of particles. \textbf{b}, Snapshots of particle probability distributions at different time points with different types of heterogeneity. Color-coding denotes values of $\rho(\boldsymbol{x}, t)$. Note that, $r(\theta; \, \boldsymbol{x})$ is symmetric in $\theta$, i.e. $r(\theta; \, \boldsymbol{x}) = r(-\theta; \, \boldsymbol{x})$, for the three types of heterogeneity, which results in distribution symmetric about the horizontal direction.}
	\end{center}
\end{figure}

In previous studies of protein diffusion and cell migration in homogeneous environments, the displacement statistics was usually analyzed by fitting the MSD to an interpolation formula from a PRW model (or Langevin equation) as $\langle \boldsymbol{x}^2 \rangle = D[t-\tau_s(1-e^{-t/\tau_s})]$, where $D$ is diffusion coefficient and $\tau_s$ is the time scale \cite{JosephSn, PeiHsun}. Here, the GPRW model provides a simple alternative Equation (Equation (5)) and its asymptotic solution (Equation (6)), which could be applied to predict protein or cell distributions in both homogeneous and heterogeneous conditions based on the moving speed and directional preference of particles.

\section{Anomalous Fickian diffusion}
Next, consider an angular heterogeneity that is correlated with the particle speed. Specifically, particles that move fast are likely to maintain their moving direction for a long time. This type of heterogeneity has been observed in several biological systems. For example, Maiuri et al. \cite{MaiuriPa} found that cells that migrate fast are likely to maintain their migratory direction because actin flows generated in motile cells reinforce cell polarity and consequently cell persistence. They showed that the relationship between persistent time $\tau_p$ and speed $U$ during a sojourn follows $\tau_p = Ae^{\lambda U}$, where $A$ and $\lambda$ are parameters obtained by fitting to experimental measurements. In the intracellular cytoplasm, active components, such as motor proteins and polymerization of the cytoskeleton, modify the fluctuating forces that act on moving particles. Since the particle speed and persistence are the result of random active forces, factors that influence forces will inevitably lead to simultaneous changes in speed and persistence. For example, when motor proteins transport a particle along a track, the external load on the particle reduces the transport speed as well as the persistence \cite{GeorgeSh, SmithDA}. Therefore, particles that frequently switch direction and particles that persist a long time can both be observed in the cytoplasm \cite{EugeneK}. 

\begin{figure}[h]
	\begin{center}
		\includegraphics[width=1.0\textwidth]{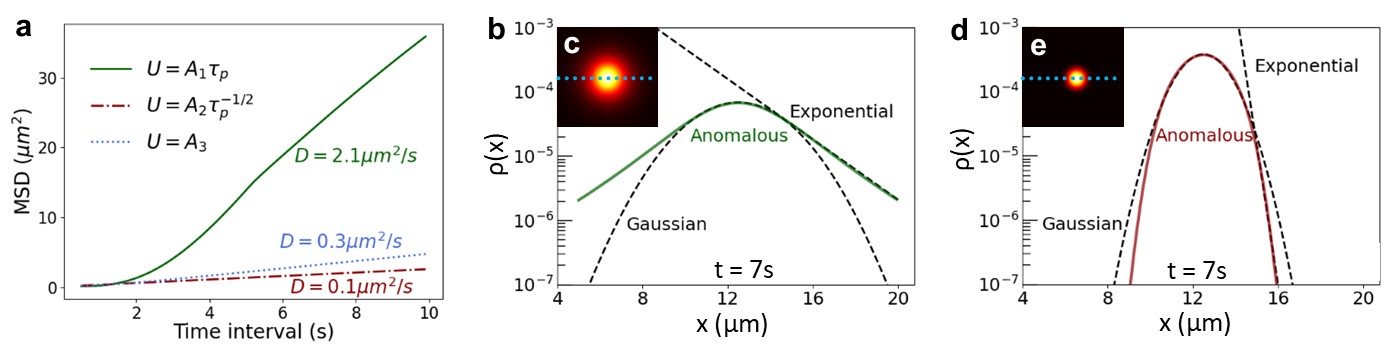}
		\caption{ \textbf{Anomalous Fickian diffusion stems from the correlated speed and persistence.} \textbf{a}, MSDs over time when the speed $U$ is positively correlated, negatively correlated and uncorrelated to the persistent time $\tau_p$. \textbf{b} and \textbf{c}, Probability distribution of particles when $t=7s$ and $U=A_1\tau_p$. \textbf{b}, Probability distribution along the center horizontal line (dashed line in \textbf{c}) shows the center part follows a Gaussian distribution and the tails follow exponential distributions. \textbf{d} and \textbf{e}, Probability distribution of particles when $t=7s$ and $U=A_2\tau_p^{-1/2}$. Color-coding in \textbf{c} and \textbf{e} denotes values of $\rho(\boldsymbol{x}, t=7s)$. $A_1=1 \mu m/s^2$, $A_2=1 \mu m/s^{\frac{1}{2}}$ and $A_3=1 \mu m/s$.}
	\end{center}
\end{figure}

To study the dynamics of the particle when its speed and persistence are correlated during each individual sojourn, particle speed $U$ is assumed to be proportional to persistent time $\tau_p$ or $\tau_p^{-1/2}$. Specifically, let $U = A_1\tau_p$ or $U=A_2\tau_p^{-1/2}$, where $A_1$ and $A_2$ are constants. Note that, when $U=A_1\tau_p$, $A_1$ might be interpreted as the instantaneous acceleration of the particle. For simplicity, let $r(\theta; \, \chi) = \dfrac{1}{2 \pi}$, and consider that $\tau_p$ still follows an exponential distribution $\varphi(\tau_p) = \gamma e^{-\gamma \tau_p}$. The master equations (1) and (2) can be rewritten as
\begin{align}
\begin{split}
q(\theta,\boldsymbol{x},t) & =  p_0 (\theta, \boldsymbol{x}) \delta (t) + \dfrac{\gamma}{2 \pi} \int_0^t q \big( A_i(t-t')\boldsymbol{n}_{\theta'},\, \boldsymbol{x}-A_i(t-t')^k\boldsymbol{n}_{\theta'}, \, t' \big) \, e^{-\gamma(t-t')} \, dt', \\
\end{split}
\\
& p(\theta, \boldsymbol{x}, t)  = \int_0^t q \big( A_i(t-t')\boldsymbol{n}_{\theta}, \, \boldsymbol{x}-A_i(t-t')^k\boldsymbol{n}_{\theta}, \, t' \big) \, e^{-\gamma (t-t')}  dt',
\end{align}

\noindent where $k=2$ and $i=1$ when $U=A_1\tau_p$, $k=\dfrac{1}{2}$ and $i=2$ when $U=A_2\tau_p^{-1/2}$, and $k=1$ and $i=3$ when $U=A_3$ is a constant. $\boldsymbol{n}_{\theta}$ is a unit vector pointing toward the $\theta$ direction. 

By numerically solving Equations (7) and (8) with periodic boundary conditions, the results show that the time required for transition to equilibrium (time scale of the system) has been extended when the speed is positively correlated with persistence (solid curve in Figure 4 a), compared with the other two cases (dashed curves). At long times, the particle motion remains Fickian diffusion (i.e. $\big<\mathbf{x}^2 \big> \propto t$). When the speed is positively correlated to persistence, the diffusion coefficient is strikingly larger than in Gaussian Fickian diffusion. Because the diffusion coefficient only holds information of second-order cumulants, we further analyzed the full probability distribution of particle displacements. When the speed and persistence are positively correlated (Figure 4 b and c), the center portion of the distribution remains Gaussian, but the probability of large displacements decays exponentially. The exponential distribution at tails stems from the self-reinforced particle persistent direction, which leads to a significant increase in the diffusion coefficient. When the speed is negatively correlated to the persistence (Figure 4 d and e), the large displacements decay exponentially faster than the Gaussian distribution. It should also be noted that the anomalous particle distribution eventually recovers to a Gaussian as required by the center limit theorem.

The non-Gaussian distribution in anomalous Fickian diffusion (also referred to as anomalous \textit{yet} Brownian diffusion by Wang et al. \cite{BoWang, BoWang2}) has been observed in several physical systems, including beads on the surface of a lipid bilayer tube, beads in an entangled solution of actin filaments, liposomes in a nematic solution and cations and anions in some ionic liquids \cite{BoWang, BoWang2, Ralf2, Zhonghan}. Granick's group and Chubynsky et al. \cite{BoWang2, Mykyta} analyzed this strange diffusion phenomenon by decomposing the particle displacement distribution into normal modes, and representing each mode by an effective diffusivity. They found that the environmental heterogeneity can be reflected by the distribution of the effective diffusivities. Here, from a physical point of view, our results indicate that a fundamental cause of the unevenly distributed diffusivities is the correlation between particle speed and persistence when the switching of direction at the end of a sojourn depends on the historical traveling speed. Further experiments can test our hypothesis that particles in the cytoplasm or other systems with the universal velocity-persistence correlation will show Fickian but non-Gaussian diffusion.

\section{Anomalous diffusion}
Next, the effects of self-reinforced directionality influenced by historical traveling distance are studied based on the GPRW model. Recall that in a memoryless system, the direction of each sojourn is independent of the history, and the persistent time is exponentially distributed. However, in many biological and ecological systems, the distribution of the persistent time is observed to be long-tailed \cite{ArielG, RheeI, KejiaChen}. A possible view of the long-tailed distribution is that the directionality is reinforced at the end of a sojourn. Particles that have performed a long sojourn are likely to maintain the traveling direction, and thus they may move a much longer distance.
\begin{figure}[h]
	\begin{center}
		\includegraphics[width=1.0\textwidth]{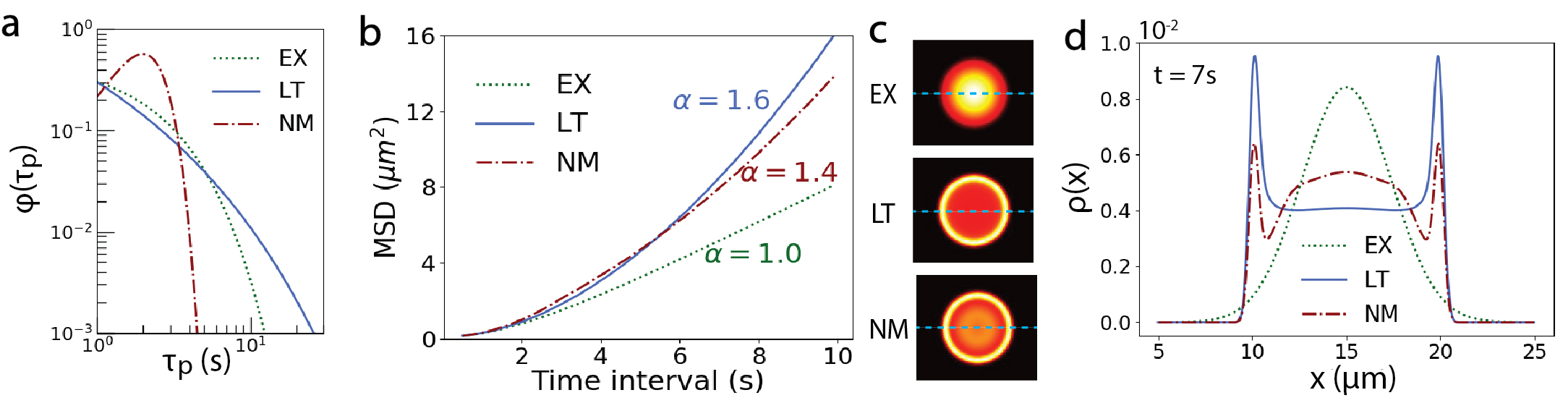}
		\caption{\textbf{The self-reinforced directionality leads to superdiffusion of particle.} \textbf{a}, The distribution of persistent time. EX represents the exponential distribution ($\varphi_{EX}(\tau_p) = \gamma e^{-\gamma \tau_p}, \, \gamma = 0.5$); LT represents the truncated long tailed distribution ($\varphi_{LT}(\tau_p) = \gamma \tau_p^{\gamma -1}e^{-\tau_p^{\gamma}}, \, \tau_p \leq 3s, \, \gamma = 0.5$); NM represents the truncated normal distribution ($\varphi_{NM}(\tau_p) = \dfrac{1}{\sqrt{2 \pi}} e^{-\frac{1}{2}(\tau_p-2)^2}, \, \tau_p \leq 3s$). \textbf{b}, The MSDs over time. \textbf{c}, Particle distributions in 2D space. Color-coding denotes values of $\rho(\boldsymbol{x}, t=7s)$. \textbf{d}, Particle distribution at the center horizontal line of the 2D space (dashed lines in \textbf{c}). }
	\end{center}
\end{figure}

In the GPRW model, for simplicity, set $U$ as a constant and $r(\theta; \, \chi) = \dfrac{1}{2 \pi}$. To study the particle dynamics with self-reinforced directionality, two types of persistence distributions other than an exponential distribution are used. Specifically, the truncated long tail distribution ($\varphi_{LT} (\tau_p) = \gamma \tau_p^{\gamma -1}e^{-\tau_p^{\gamma}}, \, \tau_p \leq 3s$) and truncated normal distribution ($\varphi_{NM}(\tau_p) = \dfrac{1}{\sqrt{2 \pi}} e^{-\frac{1}{2}(\tau_p-2)^2}, \, \tau_p \leq 3s$) (Figure 5 a). Figure 5 b shows that the long tail distribution and the normal distribution of persistence lead to superdiffusion at both short and long times. In addition, when the experimentally measured distribution $\varphi_{en} (\tau_p)$ of endosomal proteins in intracellular cytoplasm is used \cite{KejiaChen}, the results also suggest that endosomal proteins perform superdiffusion (Figure 6). Specifically, in the experiments, more than 40000 sojourns of endosomes are recorded, and the measured sojourn length was found to follow a power law distribution. By assuming the endosome moves with a constant speed, the distribution of the persistent time $\varphi_{en}(\tau_p)$ follows the same power law distribution, shown by the grey dots in Figure 6a. To calculate the endosomal distribution and its MSD over time based on the GPRW model, the distribution of persistent time is approximated by fitting the long-tailed distribution formula $\varphi_{en}(\tau_p)=\gamma \tau_p^{\gamma -1}e^{-\tau_p^{\gamma}}$ to experimental data to obtain the value of $\gamma$. When $\gamma=0.45$, the continuous approximation curve, shown by the solid curve in Figure 6 a, can be used in the master equations (1) and (2). The particle distribution and MSD are calculated by the master equations, and Figure 6 b shows that the MSD is super-linear in time, which indicates that endosomes perform superdiffusive motion.  While the majority of particles remain close to the initial position in the Gaussian Fickian diffusion, in superdiffusion, most particles move away, and form circular patterns (Figure 5 c and d). 
\begin{figure}[h]
	\begin{center}
		\includegraphics[width=0.7\textwidth]{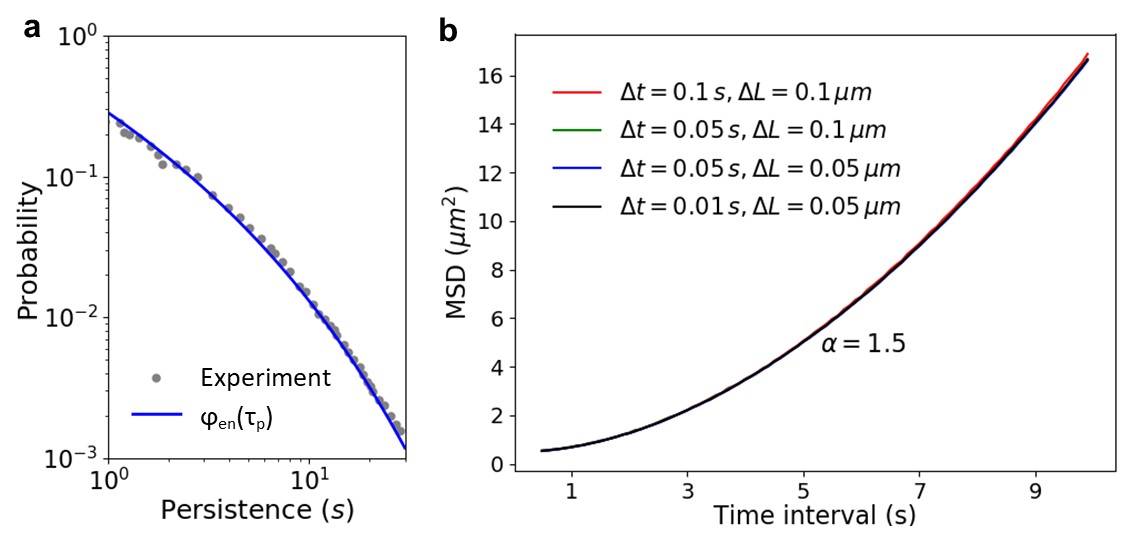}
		\caption{\textbf{The superdiffusion of intracellular protein endosomes} \textbf{a}, The experimentally measured distribution of persistence \cite{KejiaChen} of endosomes shown by grey dots, and a continuous analytic formula $\varphi_{en}(\tau_p) = 0.45 \tau_p^{-0.55} e^{-\tau_p^{0.45}}$ fitted to the experimental measurements, shown by the solid line. \textbf{b}, MSD of the particle over time calculated by using $\varphi_{en} (\tau_p)$ in the master equations (1) and (2). Different discretizations of time $\Delta t$ and space $\Delta L$ in the numerical solution of Equation (1) and (2) lead to the same results.}
	\end{center}
\end{figure}

Note that the superdiffusion is usually described by a generalized Langevin equation (GLE) as
\begin{equation}
\displaystyle m \dfrac{d}{dt} \boldsymbol{u}(t) = -m \int_0^{t} \Gamma(t-t') \boldsymbol{u}(t') dt' + \boldsymbol{F}(t),
\end{equation}
where $\boldsymbol{u}$ is the particle velocity and $\Gamma(t)$ is the memory kernel. When $\Gamma(t) \sim t^{-\alpha}$, it can be shown that $\displaystyle \lim_{t \to \infty} \langle \boldsymbol{x}^2(t) \rangle \propto t^{\alpha}$ \cite{RafaelM}. The GLE indicates that when the viscosity decreases fast enough with the time the particle has persisted ($\alpha > 1$), the process is superdiffusive at long times. Thus, superdiffusion takes place when the probability of maintaining the traveling direction increases along a persistent path. It would be interesting to further investigate the mechanisms which are used by cells or proteins to memorize their historical motions and control their directional persistence in order to achieve superdiffusive motions.

\section{Conclusions}
In this work, we proposed a GPRW model to study the effects of angular heterogeneity on anomalous diffusion and transport. We showed that when the angular heterogeneity is spatially dependent but memoryless, analytic results can be formally derived by performing a Laplace transform, and the results indicate particles perform a drifting diffusion asymptotically. In the system with memory, the mechanism of self-reinforced directionality influenced by historical speed and persistent time drives the system away from Gaussian Fickian diffusion. Specifically, when speed and persistence are correlated, particles perform Fickian but non-Gaussian diffusion. When the persistence is long-tailed distributed, particles perform superdiffusion.  Due to the clear physical meaning of variables, the GPRW model has the advantage of being applied to analyze experimental observations and make predictions. For example, the GPRW model can potentially be applied to predict protein distributions inside networks of intracellular tracks with different topology, once the angular heterogeneity at an intersection of tracks is measured in experiments. Furthermore, the GPRW model provides a more generic and flexible framework to study particle dynamis, where the time and space in the model are no longer linked by a simple constant speed. The distorted (stretched) spatial-temporal coupling, such as the self-reinforced directionality, may be the origin of anomalous diffusion and transport phenomena. Therefore, it is also interesting to study if and how the external energy input could help reconstruct the diffusion process by modifying the spatial-temporal coupling and maintaining the system out of equilibrium.

\section{Acknowledgment}
We would like to acknowledge the very useful discussions with Professor Charles Doering, in particular for insights into methods for formulating and solving the master equations. We also want to acknowledge the financial support from MCubed 3.0 program from University of Michigan.

\clearpage
\section{Appendix}
\noindent \textbf{A. Derivation of PDE (5)} 

\noindent Given master equations as
\begin{align}
\begin{split}
q(\boldsymbol{u},\boldsymbol{x},t) & =  p_0 (\boldsymbol{u}, \boldsymbol{x}) \delta (t) + \gamma \, r(\boldsymbol{u}; \, \boldsymbol{x})  \int_{-\boldsymbol{\infty}}^{\boldsymbol{\infty}} \int_0^t q[\boldsymbol{u}',\boldsymbol{x}-\boldsymbol{u}' (t-t'), t'] \, e^{-\gamma(t-t')} \, d \boldsymbol{u}' dt', \\
\end{split}
\\
& p(\boldsymbol{u}, \boldsymbol{x}, t)  = \int_0^t q[\boldsymbol{u}, \boldsymbol{x}-\boldsymbol{u} (t-t'), t'] \, e^{-\gamma (t-t')}  dt',
\end{align}
Based on a Taylor expansion of $q$, one obtains
\begin{equation}
\displaystyle q \big( \boldsymbol{u}', \boldsymbol{x}-\boldsymbol{u}'(t-t'), t' \big)=\sum_{n=0}^{\infty} \dfrac{[-\boldsymbol{u}' \cdot \bigtriangledown]^{n}}{n!} q(\boldsymbol{u}', \boldsymbol{x}, t') (t-t')^{n},
\end{equation}
Combining Equations (10) and (12) and performing a Laplace transform results in 
\begin{equation}
\displaystyle \hat{q}(\boldsymbol{u}, \boldsymbol{x}, s) = \hat{p}_0 (\boldsymbol{u}, \boldsymbol{x}) + \int_{-\boldsymbol{\infty}}^{\boldsymbol{\infty}} r(\boldsymbol{u}; \mathbf{x}) \dfrac{\gamma \hat{q}(\boldsymbol{u}', \boldsymbol{x}, s)}{s+\gamma+\boldsymbol{u}' \cdot \bigtriangledown} d\boldsymbol{u}',
\end{equation}
Performing a Taylor expansion and a Laplace transform on Equation (11) results in
\begin{equation}
\displaystyle \hat{p}(\boldsymbol{u}, \boldsymbol{x}, s) = \dfrac{\hat{q}(\boldsymbol{u}, \boldsymbol{x}, s)}{s+\gamma+\boldsymbol{u} \cdot \bigtriangledown},
\end{equation}
Combining Equations (13) and (14) results in
\begin{equation}
\displaystyle s\hat{p}(\boldsymbol{u}, \boldsymbol{x}, s) - \hat{p}_0(\boldsymbol{u}, \boldsymbol{x}) + \boldsymbol{u} \cdot \bigtriangledown \hat{p}(\boldsymbol{u}, \boldsymbol{x}, s) = \gamma \int_{-\boldsymbol{\infty}}^{\boldsymbol{\infty}} r(\boldsymbol{u}; \boldsymbol{x}) \hat{p}(\boldsymbol{u}', \boldsymbol{x}, s) d\boldsymbol{u}^{'} - \gamma \hat{p}(\boldsymbol{u}, \boldsymbol{x}, s),
\end{equation}
Performing an inverse Laplace transform leads to
\begin{equation}
\displaystyle (\partial_t + \boldsymbol{u} \cdot \boldsymbol{\bigtriangledown}) p(\boldsymbol{u}, \boldsymbol{x}, t) = \gamma \, r(\boldsymbol{u}; \boldsymbol{x}) \, \int_{-\boldsymbol{\infty}}^{\boldsymbol{\infty}} p(\boldsymbol{u}', \boldsymbol{x}, t) d \boldsymbol{u}' - \gamma p(\boldsymbol{u}, \boldsymbol{x}, t).
\end{equation}
\vspace{0.3cm}

\noindent \textbf{B. Asymptotic solution} \\
\noindent Performing a moment expansion on Equation (5) results in
\begin{align}
\displaystyle &\partial_t p_x + \dfrac{U}{2} \bigtriangledown \cdot [\overline{p}_x+\rho, \, \overline{p}_y] = \gamma \rho \int_0^{2\pi} r(\theta; \boldsymbol{x}) \cos \theta d \theta - \gamma p_x, \\
\displaystyle &\partial_t p_y + \dfrac{U}{2} \bigtriangledown \cdot [\overline{p}_y, \, \rho-\overline{p}_x] = \gamma \rho \int_0^{2\pi} r(\theta; \boldsymbol{x}) \sin \theta d \theta - \gamma p_y, \\
\displaystyle &\partial_t \rho + U \bigtriangledown \cdot [p_x, \, p_y] = 0,
\end{align}
where $p = p(\theta, \boldsymbol{x}, t)$, $p_x = p cos \theta$, $p_y = p sin \theta$, $\overline{p}_x = \int_0^{2\pi} p \cos 2 \theta d \theta$, $\overline{p}_y = \int_0^{2\pi} p \sin 2 \theta d \theta$, and $\rho = \int_0^{2\pi} p d \theta$. At long times, $\overline{p}_x = \overline{p}_y = \partial_t p_x = \partial_t p_y = 0$  \cite{Oleksandr}. Therefore, at long times, the particle motion follows
\begin{equation}
\displaystyle \partial_t \rho = \dfrac{U^2}{2\gamma} \Delta \rho - U \,  \rho \bigtriangledown \cdot \bigg[ \int_0^{2\pi} r(\theta; \boldsymbol{x}) \cos\theta d \theta, \, \int_0^{2\pi} r(\theta; \boldsymbol{x}) \sin \theta d\theta \bigg].
\end{equation}

\clearpage
\setlength{\parindent}{0in}
\bibliography{MyBib}
\end{document}